\documentclass[lettersize,journal]{IEEEtran}
\usepackage{amsmath,amsfonts}
\usepackage{algorithmic}
\usepackage{algorithm}
\usepackage{array}
\usepackage[caption=false,font=normalsize,labelfont=sf,textfont=sf]{subfig}
\usepackage{textcomp}
\usepackage{stfloats}
\usepackage{url}
\usepackage{verbatim}
\usepackage{graphicx}
\usepackage{cite}
\usepackage{color, colortbl}			
\usepackage{hyperref}
\usepackage{caption}
\usepackage{multirow, array}
\usepackage{amssymb}
\usepackage{comment}
\usepackage{xcolor}
\usepackage{tikz}
\usepackage{pgfplots}
\usepgfplotslibrary{smithchart}
\pgfplotsset{compat=1.15}
\usepgfplotslibrary{dateplot}
\pgfplotsset{grid style={white!90!black}}
\definecolor{myblue}{rgb}{0.3, 0.3, 1.0}
\definecolor{myred}{rgb}{1.0, 0.3, 0.3}
\definecolor{mygreen}{rgb}{0.0, 0.7, 0.0}
\definecolor{myyellow}{rgb}{0.9, 0.9, 0.0}
\usepackage{float}
\usepackage{soul}
\usepackage{wrapfig}

\begin{document}

\title{Implementation and Performance Analysis of a Low Resolution OFDM System Prototype With Low Cost Hardware}

\author{Eder O. de Souza, João T. Dias, Demerson N. Gonçalves
\thanks{Eder O. de Souza (ORCID: 0009-0001-8733-2854; e-mail: eder.souza@aluno.cefet-rj.br), João T. Dias (ORCID: 0000-0001-9391-9830; e-mail: joao.dias@cefet-rj.br) and Demerson N. Gonçalves(ORCID: 0000-0001-9130-7363; e-mail: demerson.goncalves@cefet-rj.br) are with Federal Center for Technological Education of Rio de
Janeiro (CEFET/RJ).}
\thanks{Digital Object Identifier: 10.14209/jcis.2023.17}}
%
\setcounter{page}{149}
\markboth{JOURNAL OF COMMUNICATION AND INFORMATION SYSTEMS, VOL. 38, No.1, 2023.}{}
%

\maketitle

\begin{abstract}
The present work focus on the implementation
and analyze of performance of a low-resolution OFDM system prototype with low-cost hardware.
A software defined radio (SDR) system was chosen in this implementation due to its various advantages over a traditional radio system.
Among the options of SDR devices available, the use of universal software radio peripherals (USRP) was avoided due to its high cost, despite its popularity in this field of research.
%
%
%
Alternatively, a combination of two low-cost SDRs, ``Hackrf One" and ``RTL-SDR Blog V3" with the GNU Radio, a popular, free and open source radio software, were used.
%
%
Thus, it was possible to emulate the behavior of a low resolution ADC in the receiver, characterize its performance and estimate its energy savings. This allowed us to determine the feasibility of building a component with the analog-to-digital conversion function with few bits of resolution. 
%
%
We conclude that the performance of an ADC with at least 5 bits of resolution is pretty reasonable and that this reduction in the number of bits, in comparison to 8-bit ADC, represents a fairly expressive energy saving.
%
%
%
%
%
%
%
%
\end{abstract}


\section{Introduction} 
\IEEEPARstart{N}{ew} applications of low latency, high connection density and ubiquitous connectivity are present in various services of 5G networks, among them we can highlight the internet of things (IoT), autonomous vehicles, telemedicine and industry $4.0$. These applications require an efficient and flexible communication system, since the characteristics of the physical (PHY) and medium access (MAC) layers in a network architecture model (like OSI and TCP/IP) affect the overall performance of the system in terms of energy efficiency, spectral efficiency, achievable throughput, quality of service (QoS), etc. \cite{mehmood2017}.
\par
An important factor in regard to energy efficiency aspect is the analog-to-digital converter (ADC) resolution, once the energy consumption is proportional to the number of bits \cite{mezghani2010}.
%
%
%
%
In this sense, some works have been take into account the power consumption mitigation provided by low resolution ADCs in potential 5G technologies \cite{mo2018}, \cite{zhang2018}.
In Reference~\cite{mo2018}, an algorithm for channel estimation in broadband millimeter wave multiple input multiple output (MIMO) systems with few-bits ADCs is proposed and simulation results are presented, while in \cite{zhang2018}, important issues of systems relying on low-resolution ADCs are discussed.
Other works are dedicated to show practical results by means of prototyping, usually with the use of  SDRs.
%
Both References \cite{gaber2021} and \cite{diouf2019} are examples of such strategy, while \cite{gaber2021} introduces a architecture that can interface different antennas to a SDR, \cite{diouf2019} proposes a SDR platform for transmission and acquisition of wideband signals.
In Reference~\cite{nervis}, the authors implemented the IEEE 802.11ah (Wi-Fi HaLow) standard, an extension of the Wi-Fi protocol focused on providing IoT-like connectivity on midrange sites (up to 1 km), simulation and experimental measurements towards evaluating the performance and limitations in interference and noise environments. 
Although these works show practical results, they lack the addressing of low resolution ADCs.
\par
The flexibility offered by SDRs make them capable of prototyping very diverse communication technologies.
Their category may range from low complexity/low cost to very advanced and expensive equipment.
Perhaps the most popular ones, in terms of paper publications, are the USRP models from National Instruments/Ettus Research.
Conversely, a low cost alternative is the HackRF One, which is popular among hobbyists but with very fewer related publications.
The References \cite{chen2021} and \cite{ishkaev2018} are some examples of works based on this device, they address long range (LoRa) collision decoding and the peak-to-average power ratio (PAPR) reduction, respectively.
However, once again, the ADC issue is not considered.
\par
Finally, \cite{wang2019} focuses on both algorithm design and system implementation in the field of low-resolution quantization communication.
The work reports the construction of a proof-of-concept prototyping system used to conduct over-the-air (OTA) tests, but, overall, expensive hardware is applied.
A summary of the topics discussed is shown in Table \ref{tab:comparison}, where is notable that none of them meet the combination of prototyping, low cost hardware and low resolution ADC.
That combination of factors is exactly what this work proposes, by the implementation of a prototyping system in order to emulate and analyse the performance of a low resolution ADC.
\begin{table}[h!]
\centering
\caption{Works comparison summary.}
\resizebox{\columnwidth}{!}{%
\begin{tabular}{|c|c|c|c|c|}
\hline
\textbf{Work}   & \textbf{Simulation} & \textbf{Prototyping} & \textbf{Low cost hardware} & \textbf{Low resolution ADC} \\ \hline
Mo et al. \cite{mo2018}       & Yes                 & No                   & No                         & Yes                         \\ \hline
Zhang et al. \cite{zhang2018}    & Yes                 & No                   & No                         & Yes                         \\ \hline
Gaber et al. \cite{gaber2021}    & No                  & Yes                  & No                         & No                          \\ \hline
Diouf et al. \cite{diouf2019}    & No                  & Yes                  & No                         & No                          \\ \hline
Nervis et al. \cite{nervis}     & Yes                  & Yes                  & No                         & No                        \\ \hline
Chen et al. \cite{chen2021}   & No                  & Yes                  & Yes                        & No                          \\ \hline
Ishkaev et al. \cite{ishkaev2018}  & No                  & Yes                  & Yes                        & No                          \\ \hline
Wang et al. \cite{wang2019}     & No                  & Yes                  & No                         & Yes                         \\ \hline
Proposed system & Yes                  & Yes                  & Yes                        & Yes                         \\ \hline
\end{tabular}%
}
\label{tab:comparison}
\end{table}
%
%
%

\par
Preliminary studies of this research were presented in References \cite{gabrielSBrT2020} and \cite{ederSBrT2022}.
In this work, we show the feasibility 
of implementing an actual low-resolution ADC and also supports further investigations in future works.
The rest of the work is divided as follows: the OFDM communication system model is presented in Section \ref{OFDM};
Section \ref{prototyping} presents the development of this model and a brief description of the implemented quantizer;
in Section \ref{results}, some performance results of the prototyping performed are presented;
in Section \ref{conclusion} the conclusions of the work are made.
%
%
%
\section{Q-OFDM} 
\label{OFDM}     
\begin{figure}[h!]
    \centering
    \includegraphics[width=.48\textwidth]{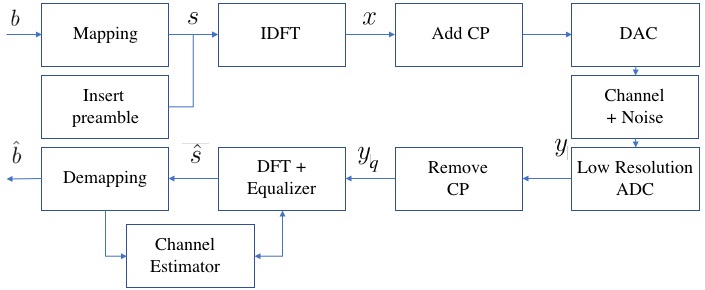}
    \caption{Block diagram of the proposed Q-OFDM system.}
    \label{fig:diagrama}
\end{figure}

%
The block diagram of the proposed quantized OFDM (Q-OFDM) system is shown in Fig.\ref{fig:diagrama}. In this system, $b$ is the bit to be transmitted, $s$ is the frequency domain data symbols, $x$ is the time domain data samples, $y$ is the time domain received signal, $y_q$ is the quantized signal , $\hat{s}$ is the received signal in the frequency domain and $\hat{b}$ are the estimated bits.
\par
The preamble is the size of an OFDM symbol and it is inserted before the data sequence. The receiver knows the preamble and uses this information to synchronize the beginning of the data symbols, correct phase and frequency deviations, as well as perform the first channel estimation.
The OFDM is a multicarrier modulation scheme based on dividing the serial data stream into parallel groups of streams that are transmitted on orthogonal subcarriers.
When an OFDM signal crosses a time-dispersive or frequency-selective channel, the subcarriers are only affected by a constant or frequency flat channel.
In this type of channel, the signal can be distorted by intersymbolic interference (ISI).
This problem is solved in OFDM by inserting a copy of the last part of the symbol, also known as a cyclic prefix (CP), at the beginning to absorb the channel delay spreads.
The OFDM signal can be expressed in the time domain by \cite{ref3}
\begin{equation}
    \it{x} [n]=\displaystyle \sum_{k=0}^{K-1} \it{s}_{k}e^{j2\pi \frac{ k}{K}n},
    \label{eq:ofdmtransmit}
\end{equation}
where $\it{s}_{k}$ is the data symbol on the $k$-th subcarrier and $K$ is the number of subcarriers in the OFDM symbol.

The signal at the receiver input can be written by
\begin{equation}
    \it{\mathbf{y}} = \it{\mathbf{x}}*\it{\mathbf{h}}+\it{\mathbf{\omega}},
 \label{eq:ofdmreceived}
\end{equation}
where $\it{\mathbf{y}} \in \mathbb{C}^{(K+CP+N_p-1) \times 1}$, $CP$ is the length of the cyclic prefix, and $N_p$ is the number of paths considered in the channel, $\it{\mathbf{x}} \in \mathbb{C}^{(K+CP) \times 1}$, $\it{\mathbf{h}} \in \mathbb{C}^{(N_p)\times (1)}$ is the channel impulse response, $*$ is the convolution operation, and $\it{\mathbf{\omega}} \in \mathbb{C }^{(K+CP+N_p-1)\times 1}$ is additive white Gaussian noise (AWGN).
%
%
%
\subsection{Low resolution ADC}
\label{subsec:II.A}
\begin{figure}[h!]
    \centering
    \includegraphics[width=1\columnwidth]{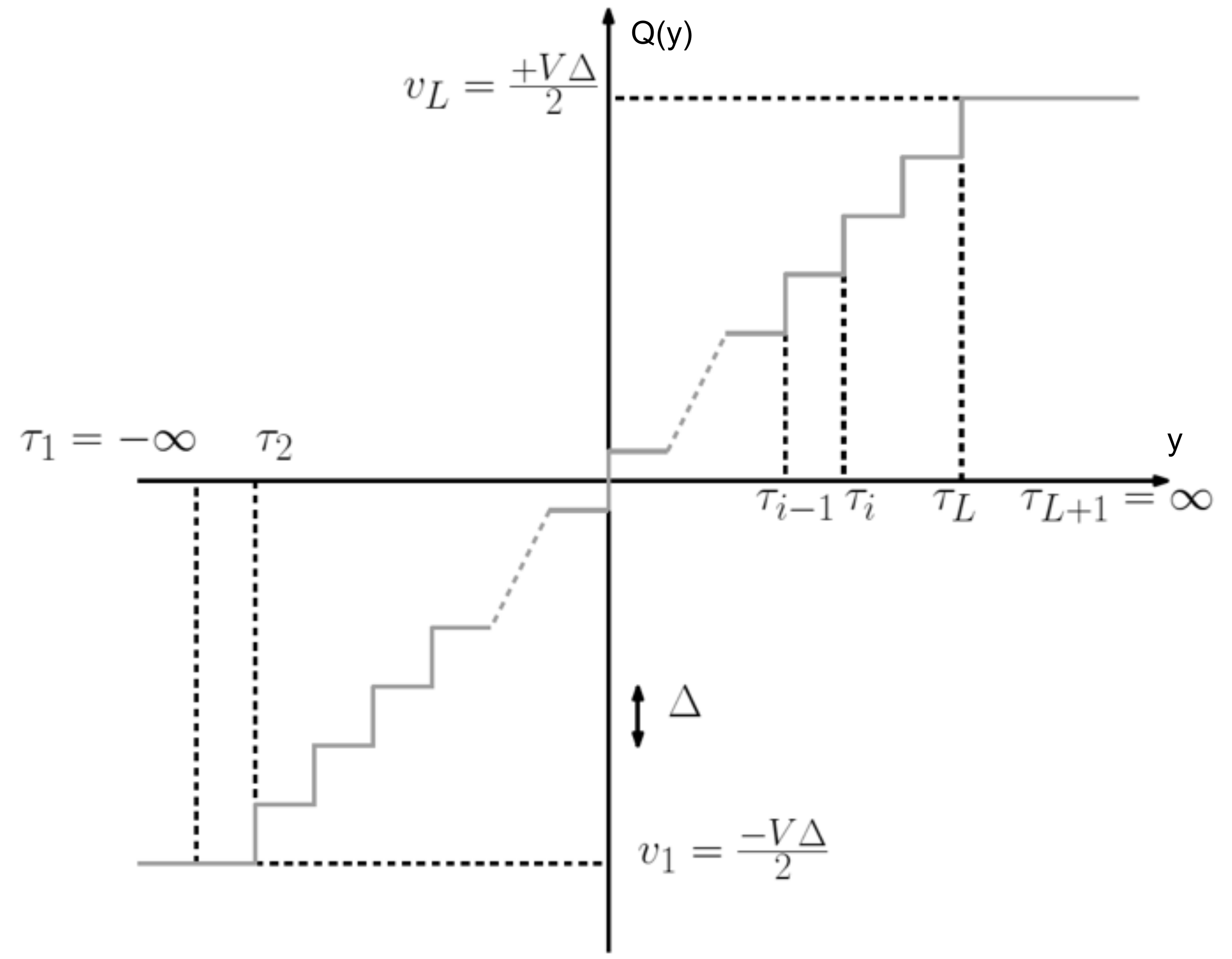}
    \caption{Characterization of the uniform and symmetrical \textit{Mid-Riser} quantizer.}
    \label{Fig2}
\end{figure}

%
In this section, the low resolution ADC considered in this work and its main characteristics will be described.
%
The choice of the linear quantizer \textit{Mid-Riser} is due to its wide use, easy implementation and good performance \cite{ref6}.
This quantizer divides the signal to be mapped into equidistant quantization levels, with a step size $\Delta$ and total number of quantization levels ($V$) given by $V=2^{\flat}$, where $\flat$ is the number of quantization bits used in the ADC.
For uniform and symmetric \textit{Mid-Riser} quantizers, the quantizer output levels are given by the equation
\begin{equation} 
    v_i=\frac{-V\Delta}{2}+(i-\frac{1}{2})\Delta.
\end{equation}
The quantizer input limits are defined by $\tau_1=-\infty$, $\tau_{L+1}=\infty$ and $\tau_i=v_i+\frac{\Delta}{2}$, for $i = 2,3,...,V$ \cite{ref6}.
Therefore, for a discrete input signal $y$, the characterization of this quantizer is given by the Eq. (~\ref{eq:adc}) and illustrated in Fig. \ref{Fig2}.
%
%
\begin{equation}
    Q(y[n])=
    \begin{cases}
        v_1, & y \leq \tau_1, \\
        v_i, & \tau_{i-1} < y \leq\tau_i, \\
        v_L, & y > \tau_{L+1}.
    \end{cases}
    \label{eq:adc}
\end{equation}

The signal after the quantizer will take the form
\begin{equation}
    \it{\mathbf{y}}_{q}= \mathcal{Q}_c(\it{\mathbf{y}}),
 \label{eq:signalquant1}
\end{equation}
where $\mathcal{Q}_c(.)$ denotes the elementary complex-valued quantizer mapping function, which comprises two parallel real-valued quantizers $\mathcal{Q}(.)$ that independently quantize the real and imaginary parts of each analog input sample \cite{ref9}.

The received signal in the frequency domain can be written as
\begin{equation}
    \it{\mathbf{\hat{s}}}= g(\mathbf{F}\it{\mathbf{y}}_{q}),
 \label{eq:signalquant2}
\end{equation}
where $\mathbf{F}$ is the discrete Fourier transform (DFT), a unitary matrix
of dimension $K\times K$ and $g(.)$ is the function of the equalizer.
%
%
%
%
%
%
%
\par
Due to the fact that the decision feedback equalizer (DFE) is native to the OFDM library in GNU Radio \cite{dfe-gnuradio} and the objective is to observe in the MatLab simulation the behavior of the bit error rate (BER) that must be found in the prototyping, the OFDM system was simulated with the same DFE equalizer present in the GNU Radio library, that is, the simple DFE (SDFE).
\par
In the SDFE, for each symbol $t$ and for each subcarrier $k$ in each symbol, if it is a pilot carrier then we update the channel state for $k$ to be
\begin{equation}
    \it{H[k]}= \alpha \it{\hat{H}[k]}+(1-\alpha)\frac{\hat{s}_{t,k}}{p_{k}},
 \label{eq:sdfe}
\end{equation} 
where $\it{H[k]}$ is the frequency response of the channel, $\alpha$ is a parameter normally set to $0.1$, $\it{\hat{H}[k]}$ is a channel response estimation obtained and applied in the forward filter, ${p_{k}}$ is the pilot symbol corresponding to the $k$th index, and $\hat{s}_{t,k}$ is the estimated received symbol.
The expression $(1-\alpha)\frac{\hat{s}_{t,k}}{p_{k}}$ is obtained and applied in the feedback filter.
%
%
%
%
\par
\begin{figure}[h!]
    \centering
    \includegraphics[width=.8\columnwidth]{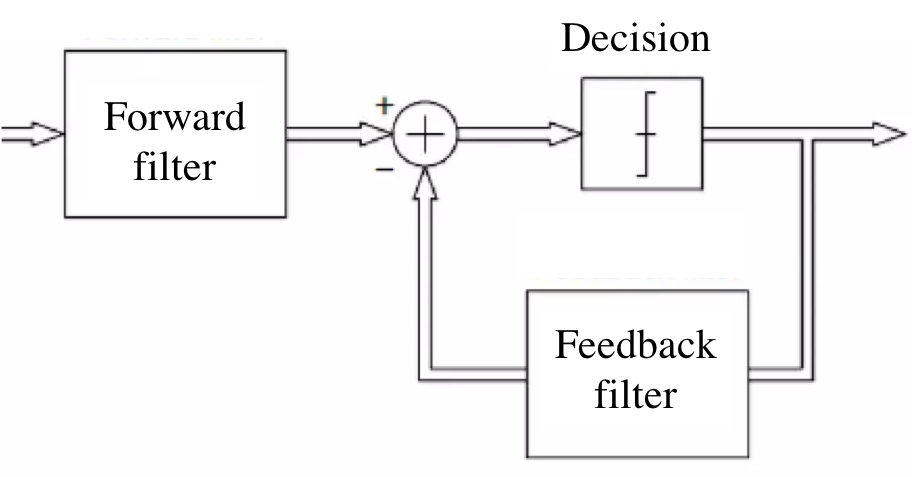}
    \caption{Simplified diagram of SDFE.}
    \label{dfe}
\end{figure}

If the subcarrier is not a pilot subcarrier, then we update the channel in a similar way. First, let the equalized symbol just be 
$\frac{\Bar{s}_{t,k}}{\it{H_t[k]}}$, where $\Bar{s}_{t,k}$ is the received symbol after the FFT and before the equalizer. 
%
Next, we use the constellation mapper to essentially “snap” the complex item to a complex item that we know.
Now that we know what this symbol really was (the decision), we can use this feedback to update our channel slightly, via the same equation above, Eq.(\ref{eq:sdfe}), just using the snapped complex number instead of the pilot symbol. The Fig. \ref{dfe} shows the simplified diagram of SDFE.
\par
We believe that the choice of GNU Radio for the DFE equalizer is due to two disadvantages for OFDM systems free of intercarrier and intersymbol interference. First, a longer guard interval than the delay channel has to use in each OFDM symbol period, thus resulting in a considerable loss in bandwidth utilization efficiency. Second, FFT-based demodulation methods, although computationally simple, do not use enough statistics for channel equalization, which degrades performance \cite{sun1999}.
%
%
%
\section{Q-OFDM system prototyping with SDRs} 
\label{prototyping}                           
%
Here we discuss the details of the implementation of our prototype system.
In order to better organize the information we divided this section into three subsections: characteristics of the hardware [\ref{subsec:III.A}], development of the low resolution ADC block [\ref{subsec:III.B}] and, finally, the description of the transmitter and receiver projects [\ref{subsec:III.C}].
%
%
%
\subsection{Hardware used}
\label{subsec:III.A}
%
\par
The SDRs devices selected for this work were the HackRF One \cite{hackrf} and the RTL-SDR Blog V3 \cite{rtl-sdr}.
Some information about these devices is shown in the Table \ref{tab:SDRs}, where we notice that there is a window of compatibility between the SDRs in relation to the values of the operating frequency and the sampling rate.
The chosen sampling rate was $1.8 \, \mbox{MS/s}$ (Mega Samples/seconds) and the chosen operating frequency was dependent on the antennas.
\begin{table}[h!]
\centering
\caption{Basic characteristics of the SDRs}
\label{tab:SDRs}
\resizebox{0.45\textwidth}{!}{%
\begin{tabular}{c|c|c|}
\cline{2-3}
\textbf{}                                                            & \textbf{HackRF One} & \textbf{RTL-SDR Blog V3}      \\ \hline
\multicolumn{1}{|c|}{\textbf{Communication}}                         & half duplex          & simplex reception             \\ \hline
\multicolumn{1}{|c|}{\textbf{Operating frequency}}                    & 1 MHz - 6 GHz        & 500 kHz - 1766 MHz            \\ \hline
\multicolumn{1}{|c|}{\multirow{2}{*}{\textbf{Converter resolution}}} & A/D - 8 bits         & \multirow{2}{*}{A/D - 8 bits} \\ \cline{2-2}
\multicolumn{1}{|c|}{}                                               & D/A - 8 bits         &                               \\ \hline
\multicolumn{1}{|c|}{\textbf{Sample rate}}                           & 20 MS/s              & 2.4 MS/s                      \\ \hline
\end{tabular}%
}
\end{table}
%
%
%

\par
The antennas used in the transmitter and receiver were both simple monopole telescopic antennas with SMA connector.
In the case of HackRF One, which acted as the transmitter, in addition to the antenna, a DC-Blocker was also used, in order to provide a purely AC signal to the antenna.
\par
The performance of the chosen antenna was characterized in the range from $350$ to $450 \, \mbox{MHz}$ using the NanoVNA-H Rev3.6 \cite{nanovna}, a cheap Vector Network Analyzer.
And by using the NanoVNASaver \cite{nanosaver}, a software tool for the NanoVNA, we saved a Touchstone file, analyzed the data and plotted the Smith chart and the VSWR graph, shown in the Fig. \ref{fig:smith_chart} and Fig. \ref{fig:vswr}, respectively.
Based on this analysis we decided that $410 \, \mbox{MHz}$ was good operating frequency for our prototype, because in this frequency the antenna yields the minimum VSWR value ($1.238$) and has an impedance that is close enough to the input impedance of both SDR devices ($50 \, \Omega$).
\begin{figure}[h!]
    \centering
    \begin{tikzpicture}
        \begin{smithchart}[
            title=Antenna Impedance,
            show origin,
            ]
            \addplot coordinates {(1.16976907, -0.15694837)};
            \addplot [blue] file {data/impedance.dat};
            \node[label={180:{\textcolor{blue}{$1.17 - j0.157$}}}] at (axis cs:1.16976907, -0.3) {};
            \node[label={180:{\textcolor{blue}{@$410 \, \mbox{MHz}$}}}] at (axis cs:0.95, -0.55) {};
        \end{smithchart}
    \end{tikzpicture}
    \caption{Smith chart - antenna analysis between $350$ and $450 \, \mbox{MHz}$.}
    \label{fig:smith_chart}
\end{figure}
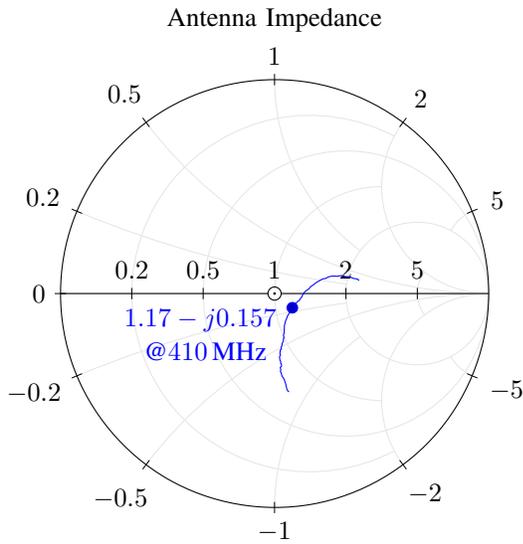
\begin{figure}[h!]
    \centering
    \begin{tikzpicture}
        \begin{axis}[
            grid=both,
            title=Antenna VSWR,
            xlabel={frequency [$\mbox{MHz}$]},
            ylabel={VSWR},
            ]
            \addplot [blue] file {data/VSWR_new.dat};
            \node[circle,fill={blue},inner sep=2pt] at (axis cs:410, 1.238) {};
            \node[anchor=west] (source) at (axis cs:408, 1.25){};
            \node (destination) at (axis cs:408, 1.7){\textcolor{blue}{$1.238$@$410 \, \mbox{MHz}$}};
            \draw[->,blue](source)--(destination);
        \end{axis}
    \end{tikzpicture}
    \caption{VSWR graph - antenna analysis between $350$ and $450 \, \mbox{MHz}$.}
    \label{fig:vswr}
\end{figure}
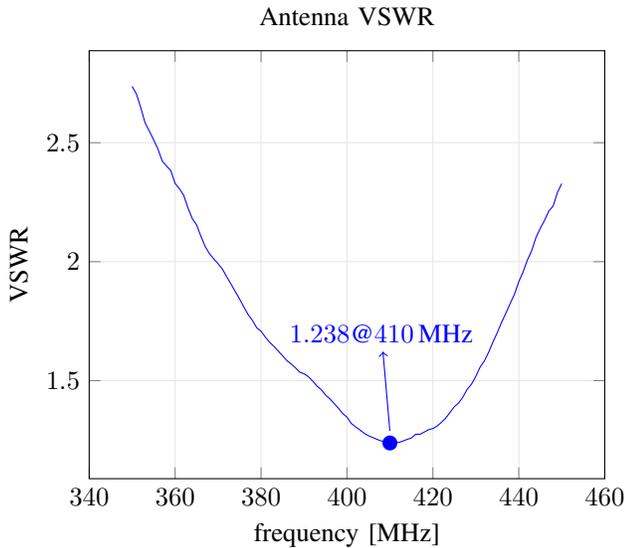
%
%
%
%
\par
The notebooks used in this work were the CCE Win T54P+, which combined with the HackRF One formed the transmitter SDR system; and the Lenovo idepad 310 - 14ISK notebook, which combined with the RTL-SDR Blog V3 formed the receiver SDR system.
The Table \ref{tab:computers} summarizes some basic technical characteristics of both computers.
The version 3.10 of GNU Radio was installed in both of them.
%
%
%
\begin{table}[h!]
    \centering
    \caption{Specifications of computers used}
    \resizebox{.45\textwidth}{!}{%
    \begin{tabular}{|cc|c|}
    \hline
    \multicolumn{2}{|c|}{\textbf{\textbf{Notebook 1}}} & \textbf{CCE Win T54P+} \\ \hline
    \multicolumn{1}{|c|}{\multirow{3}{*}{CPU}} & processor & Intel\textsuperscript{\textregistered} Core\textsuperscript{TM} i5 M 450 \\ \cline{2-3}
    \multicolumn{1}{|c|}{} & cores & 4 \\ \cline{2-3}
    \multicolumn{1}{|c|}{} & frequency & 2,4 GHz \\ \hline
    \multicolumn{2}{|c|}{RAM} & (2$\times$) 2 GB DDR3 - 1333 MHz SODIMM \\ \hline
    \multicolumn{2}{|c|}{Operating System} & Ubuntu 22.04.2 LTS \\ \hline \hline
    \multicolumn{2}{|c|}{\textbf{Notebook 2}} & \textbf{Lenovo idepad 310 - 14ISK} \\ \hline
    \multicolumn{1}{|c|}{\multirow{3}{*}{CPU}} & processor & Intel\textsuperscript{\textregistered} Core\textsuperscript{TM} i3-6100U \\ \cline{2-3 }
    \multicolumn{1}{|c|}{} & cores & 4 \\ \cline{2-3}
    \multicolumn{1}{|c|}{} & frequency & 2,3 GHz \\ \hline
    \multicolumn{2}{|c|}{RAM} & 4 GB DDR4 - 2133 MHz SODIMM \\ \hline
    \multicolumn{2}{|c|}{Operating System} & Ubuntu 22.04.2 LTS \\ \hline
    \end{tabular}}
    \label{tab:computers}
\end{table}
\par
\begin{figure}[h!]
    \centering
    \includegraphics[width=0.48\textwidth]{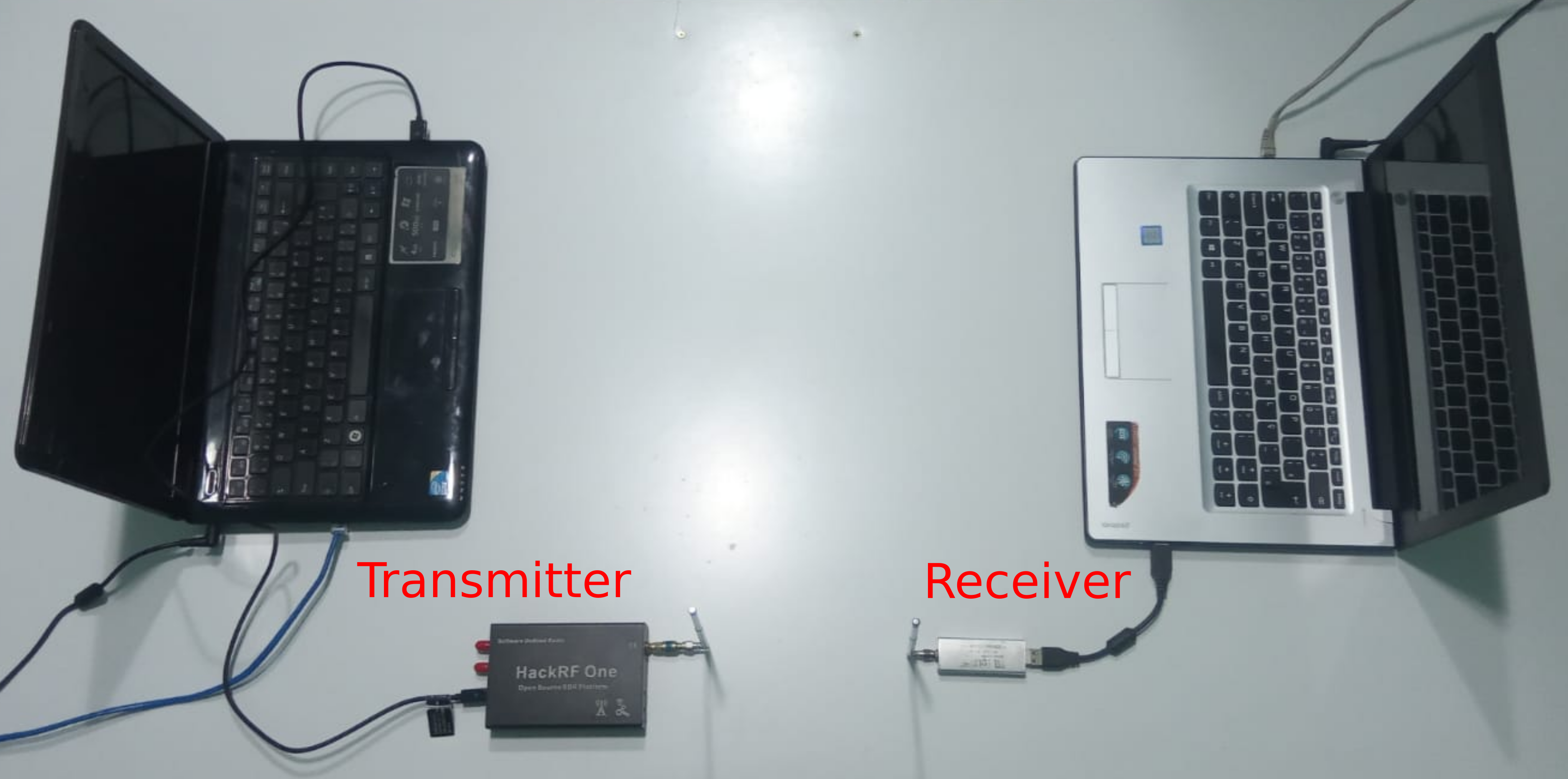}
    \caption{SDR system prototype setup.}
    \label{fig:sdrs_setup}
\end{figure}

The Fig. \ref{fig:sdrs_setup} shows the complete setup of the prototype system. The transmitter SDR was placed on the left side and the receiver SDR on the right side, with a distance of 1 m between the antennas.
%
%
%
\subsection{Low resolution ADC block}
\label{subsec:III.B}
\par
\begin{figure}[h!]
    \centering
    \includegraphics[width=0.4\textwidth]{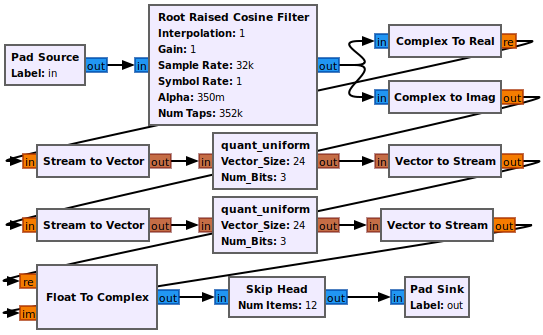}
    \caption{Low resolution ADC block diagram.}
    \label{fig:lrc}
\end{figure}

The Fig. \ref{fig:lrc} shows the GNU Radio Companion (GRC) project of the low resolution ADC block implemented by the authors.

Note that this block is actually composed by others blocks, which the GNU Radio documentation classifies as a ``Hierarchical Block'' \cite{hierblock}.
The signal flow passes through two main blocks: the ``Root Raised Cosine Filter'', which has the function of simulating a DA conversion; and the ``Quantizer'', that quantizes the signal.
The latter was created by the authors from an ``Embedded Python Block'' \cite{embblock}, a special type of GNU Radio block that allows the user to customize it through a Python script.
\par
Thus, the low resolution ADC block was available locally in the GRC library under the name of ``Low Resolution Converter'' and was used in the receiver design with different resolution bits.
%
%
%
\subsection{Transmitter and receiver designs}
\label{subsec:III.C}
\begin{figure*}[h!]
    \centering
    \includegraphics[width=.7\textwidth]{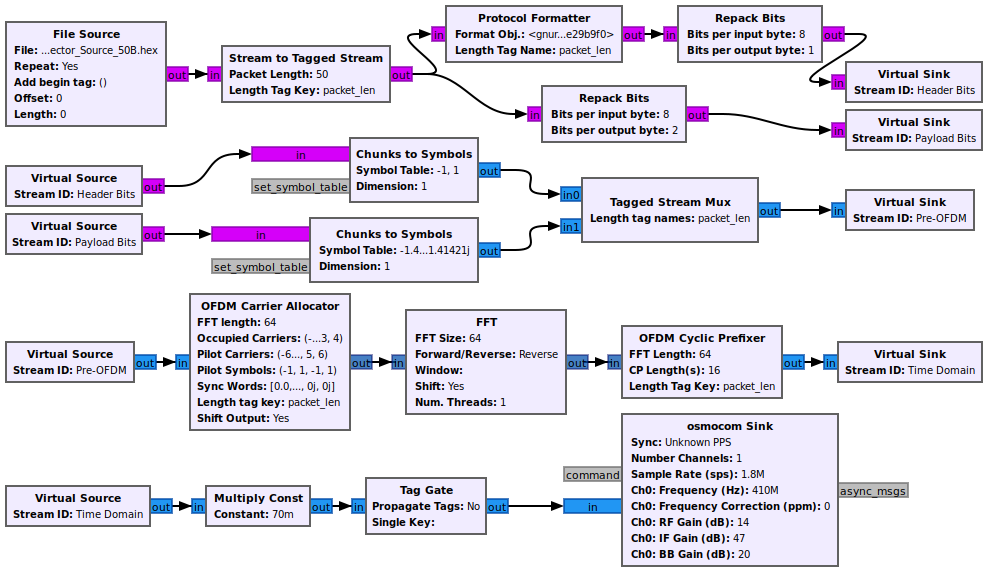}
    \caption{Block diagram of transmitter.}
    \label{fig:OFDM_Tx_GRC}
\end{figure*}

%
%
%
The GNU Radio code and projects used in this work are publicly available on the GitHub platform \cite{github}.
The transmitter design is shown in Fig. \ref{fig:OFDM_Tx_GRC}.
Specific blocks are used for each stage of an OFDM transmission and, on purpose, the resource of an error correcting code is not used.
The interface with HackRF One is done through the ``osmocom Sink'' block.
%
%

The transmission consisted of sending 10 randomly generated binary files of $50 \, \mbox{B}$ each for every noise level increment at the receiver side.
In the reception, each received file had a size of $15 \, \mbox{kB}$, which means $300$ repetitions of the transmitted file. To calculate the BER of each noise level we first calculated the BER related to each of the $10$ files of that specific level and then we averaged the results. This process, repeated at various noise levels.
\par
\begin{figure*}[h!]
\centering
\includegraphics[width=1\textwidth]{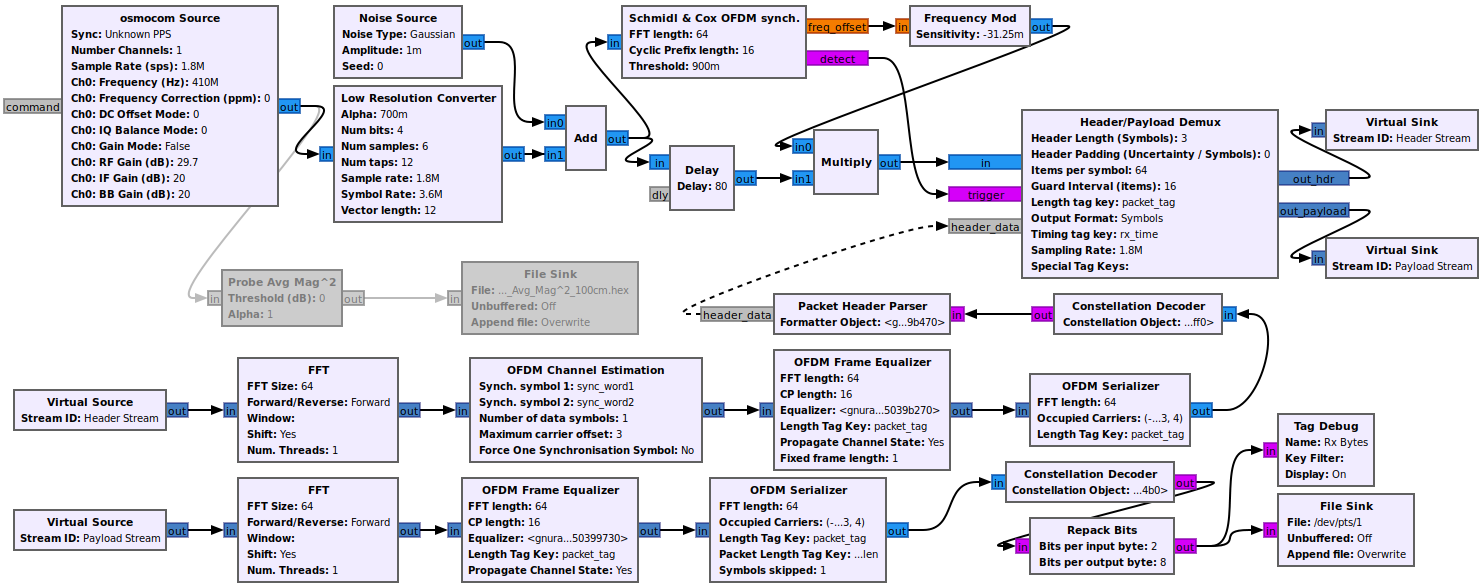}
\caption{Block diagram of receiver.}
\label{fig:OFDM_Rx_GRC}
\end{figure*}

The Fig. \ref{fig:OFDM_Rx_GRC} shows the receiver design and configuration that allows the collection of data from three distinct events: the average signal power, the low resolution ADC performance estimation and the receiver SDR performance.
The information regarding the average signal power is given by the ``Probe Avg Mag\^~2'' block \cite{probe_block}.  
The ``Low Resolution Converter'' block emulates the behavior of an ADC with variable bit resolution.
By disabling it from the project it is possible to collect data directly from the RTL-SDR's 8-bit ADC.
Similar to the transmitter design, specific blocks are used for each stage of an OFDM reception with the omission of an error correction and the interface with the SDR done by the ``osmocom Source'' block.
Furthermore, there was also the addition of the Gaussian noise, through the ``Noise Source'' block, to provide different levels of signal to noise ratio (SNR).
\par
%
Thus, knowing the noise levels and the average power of the signal it was possible to calculate the SNR and the survey of the BER curves shown in Section \ref{results}.
The measurements were performed with a distance of 1 meter between the antennas.
%
%
%
\section{Results} 
\label{results}   
%
First, to evaluate the impact of the number of quantizer bits on the power consumption of the ADC, the power consumption curve per number of quantization bits was drawn up. To this end, the expression described in \cite{ref16} was used, which lists the energy consumed by an ADC in relation to the number of resolution bits:
\begin{equation}
    P_{ADC}=cf_s2^{\flat},
\end{equation}
where $\flat$ represents the number of bits of resolution, $c$ is a constant associated with the noise merit figure, and $f_s$ is the sampling rate.
\par
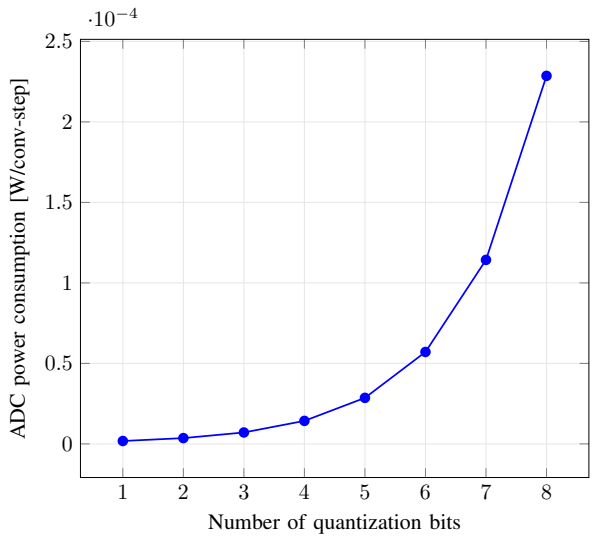
\begin{figure}[h!]
    \centering
    \resizebox{.43\textwidth}{!}{%
    \begin{tikzpicture}
        \begin{axis}[
            scale only axis,
            width=0.45\textwidth,
            grid=both,
            xlabel={Number of quantization bits},
            ylabel={ADC power consumption [W/conv-step]},
            ]
            \addplot[color=blue,mark=*,thick] table[col sep=comma, x={nbits}, y={power}] {data/power_consumption.csv};
        \end{axis}
    \end{tikzpicture}}
    \caption{ADC power consumption as a function of the number of quantizer bits.}
    \label{Fig6}
\end{figure}

This simulation was carried out with $f_s=1.8$ MHz and $c=496$ fJ/conv-step (Fento Joule / conversion-step) \cite{ref17}. As we can see in Fig. \ref{Fig6}, the impact of power consumption is more significant when the number of bits is greater than $5$ bits.

    In order to validate the proposed system and analyze its performance by varying the number of bits, BER curves were initially created by simulation in MatLab considering the parameters shown in the Table ~\ref{parameters}:
\begin{table}[h!]
    \centering
    \caption{Simulation parameters.}
    \label{parameters}
    \vspace{-.5 em}
    \begin{tabular}{|l|c|}
        \hline \hline
        number of subcarriers [K] & 64\\ \hline  
        subcarrier modulation & QPSK\\ \hline
        preamble modulation & BPSK\\ \hline 
        cyclic prefix length in number of subcarriers & 16 \\ \hline  
        number of quantization bits [$\flat$] & 1 to 8 \\ \hline
        Equalizer & SDFE \\
        \hline  \hline
    \end{tabular}
\end{table}
\par
To choose the channel model used in the simulation, all multipath channel models presented in Annex $4$ of the ITU-R BT.2035 \cite{itu} document, for the scenario in question, were analyzed and we came to the conclusion that the model that presented the characteristics closest to those we had in our laboratory was an adaptation with the first 4 taps of the "Brazil D" model, which is described in the Table ~\ref{channel}.
\begin{table}[h!]
    \centering
    \caption{Channel Brazil D.}
    \label{channel}
    \vspace{-.5 em}
    \begin{tabular}{|l|c|c|c|c|}
        \hline \hline
         Delay [$\mu$s]& $0.15$ & $0.63$ & $2.22$ & $3.05$ \\ \hline  
         Attenuation [dB] & $0.1$ & $3.8$ & $2.6$ & $1.3$ \\ \hline           
        \hline  \hline
    \end{tabular}
\end{table}
%
%
%
%
\par
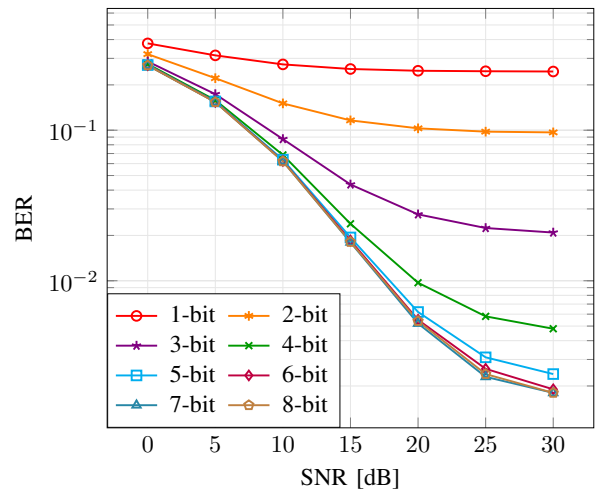
\begin{figure}[h!]
	\centering
	\resizebox{.43\textwidth}{!}{%
	\begin{tikzpicture}
	\begin{semilogyaxis}[
	scale only axis,
	width=0.39\textwidth,
	grid=both,
	xlabel={SNR [$\mbox{dB}$]},
	ylabel={BER},
	legend entries={1-bit,2-bit,3-bit,4-bit,5-bit,6-bit,7-bit,8-bit},
        legend columns=2,
	legend cell align=left,
        legend style={at={(0,0)},anchor=south west,inner sep=3pt, style={column sep=0.1cm}}
	]
	\addplot[color=red,mark=o,thick] table[col sep=comma, x={SNR}, y={1b}] {data/bf_zf.csv};
	\addplot[color=orange,mark=asterisk,thick] table[col sep=comma, x={SNR}, y={2b}] {data/bf_zf.csv};
	\addplot[color=violet,mark=star,thick] table[col sep=comma, x={SNR}, y={3b}] {data/bf_zf.csv};
	\addplot[color=green!60!black,mark=x,thick] table[col sep=comma, x={SNR}, y={4b}] {data/bf_zf.csv};
	\addplot[color=cyan,mark=square,thick] table[col sep=comma, x={SNR}, y={5b}] {data/bf_zf.csv};
	\addplot[color=purple,mark=diamond,thick] table[col sep=comma, x={SNR}, y={6b}] {data/bf_zf.csv};
	\addplot[color=cyan!60!black,mark=triangle,thick] table[col sep=comma, x={SNR}, y={7b}] {data/bf_zf.csv};
	\addplot[color=brown,mark=pentagon,thick] table[col sep=comma, x={SNR}, y={8b}] {data/bf_zf.csv};
	\end{semilogyaxis}
	\end{tikzpicture}}
	\caption{BER performance with the number of quantization bits ranging from 1 to 8.}
	\label{Fig4}
\end{figure}

The Fig. \ref{Fig4} shows the performance of the BER as a function of the SNR in the MatLab simulation.

\par
The satisfactory results of the simulation motivated the realization of practical tests with hardware, that is, with the prototype based on SDRs.
\par
\begin{figure}[h!]
    \centering
    \resizebox{.43\textwidth}{!}{%
    \begin{tikzpicture}
        \begin{semilogyaxis}[
            scale only axis,
            width=0.45\textwidth,
            grid=both,
            xlabel={SNR [$\mbox{dB}$]},
            ylabel={BER},
            legend entries={Hardware, LRC 3b,LRC 4b,LRC 5b},
            legend columns=2,
            legend cell align=left,
            legend style={at={(0.5,-0.165)},anchor=north,inner sep=3pt, style={column sep=0.15cm}}
            %
            %
            %
            ]
            \addplot[color=blue,mark=+,thick] table[col sep=comma, x={SNR}, y={Hardware}] {data/data_review.csv};
            \addplot[color=violet,mark=star,thick] table[col sep=comma, x={SNR}, y={3b}] {data/data_review.csv};
            \addplot[color=green!60!black,mark=x,thick] table[col sep=comma, x={SNR}, y={4b}] {data/data_review.csv};
            \addplot[color=cyan,mark=square,thick] table[col sep=comma, x={SNR}, y={5b}] {data/data_review.csv};
        \end{semilogyaxis}
    \end{tikzpicture}}
    \caption{Prototype system performance with $1 \, \mbox{m}$ distance between the antennas.}
    \label{fig:performanceReview_1m}
\end{figure}
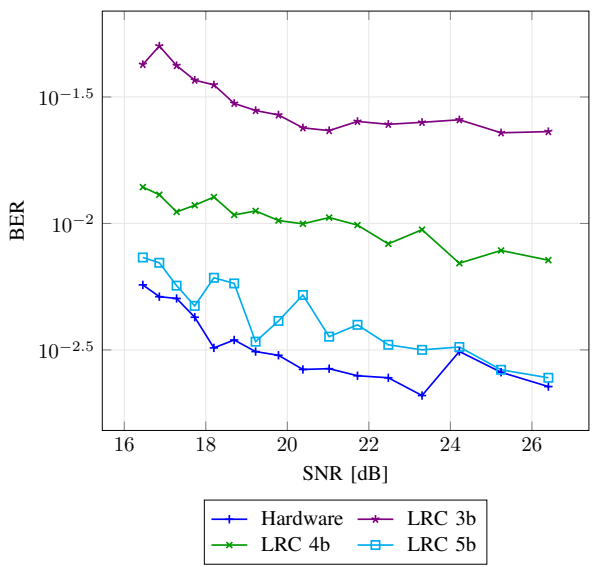

The results of our prototype system are shown in the Fig. \ref{fig:performanceReview_1m}, where the performance of the receiver SDR is indicated by the ``Hardware'' curve and the other curves are results for different resolution bits in the ``Low Resolution Converter'' block of the receiver project, which is indicated in the graph legend by the term ``LRC'' followed by the number of bits used in each measurement.
\par
Due to the fact that the 1 and 2 bit resolution curves present very poor BER performance and generate a plateau with a very high BER value during the simulation with MatLab, Fig.\ref{Fig4}, they were not considered in our tests with the prototype.
%
%
%
%
%
\par
As the MatLab simulation results show that the 6, 7 and 8 bit resolution curves present overlapping in the BER performance, Fig.\ref{Fig4}, and these cases imply greater energy consumption, as we see in fig.\ref{Fig6}, in addition to the fact that we observed overrun problems for resolutions greater than 5 bits in several tests carried out. For these reasons, we removed 6, 7 and 8 bit resolutions curves from our analysis.
\par
As described in the session \ref{subsec:III.C}, to generate the curves were sent bursts of 10 randomly generated binary files of $50 \, \mbox{B}$ each for every noise level increment at the receiver side.
In the reception side, for each received burst, an frame of size of $15 \, \mbox{kB}$, which means $300$ repetitions of the transmitted file is used to  obtain the SNR and BER calculate.
All curves of the graph in Fig. \ref{fig:performanceReview_1m} falls within the SNR range from $16.456 ~ \mbox{dB}$ to $26.403 ~ \mbox{dB}$.
The ``LRC 3b'' curve starts with a BER of $4.2498 \times 10^{-2}$ and after a SNR of $20 ~ \mbox{dB}$ it reaches a plateau around a BER of $2.5 \times 10^{-3}$.
The ``LRC 4b'' curve starts with a BER of $1.3923 \times 10^{-2}$  and we did not observe the occurrence of a plateau, in agreement with what was observed in the MatLab simulation, Fig.\ref{Fig4}.
The ``LRC 5b'' curve starts with a BER of $7.3375 \times 10^{-3}$ and presents a BER performance very close to the ``Hardware'' curve that starts with a BER of $5.7167 \times 10^{-3}$, as we expected.

%
%
%
\section{Conclusion} 
\label{conclusion}
%
Based on the prototyping work presented in this research, it is safe to say that there is a basis for creating critical hardware for systems that require high energy efficiency, that is, an ADC with less than 8 bits of resolution.
%
It is possible to reinforce the importance of the subject discussed, since it can strongly impact the prototyping of radio systems in the laboratory.
The use of SDRs allows flexibility to test different radio communication systems with the same device.
And although SDRs such as USRP are already widely used in research, this work has shown that there are low-cost devices on the market that can fulfill this function.
%
The contents presented here demonstrate that many other researches can still be carried out on prototyping using low-cost SDRs, due to the importance of the topic and numerous contributions to the academic environment.
As an example of future work, one can consider implementing a receiver in GNU Radio that takes into account the quantization noise and thus improves the performance of the presented system even when running in low resolution.
%
%
%

%
%
%

\vfill

\begin{wrapfigure}{L}{0.2\textwidth}
    \centering
    \includegraphics[width=0.2\textwidth]{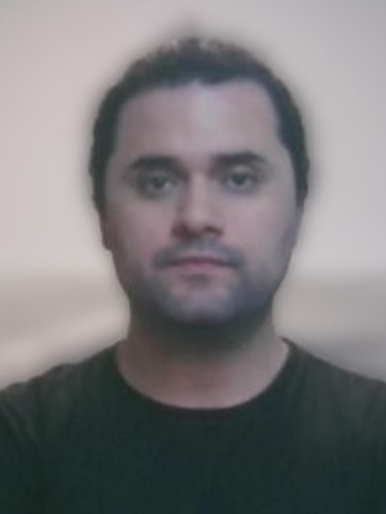}
\end{wrapfigure}

\noindent \textbf{Eder Oliveira de Souza} was born in the city of Rio de Janeiro, Brazil, on July 1st, 1990. He received his bachelor's degree in Telecomunication Engeneering from Fluminense Federal University (UFF) in 2016. In 2022, he received his bachelor's degree in Electronic Engineering from Federal Center for Technological Education of Rio de Janeiro (CEFET/RJ).  His research  interests  are digital signal processing and digital transmission systems. \newline

\begin{wrapfigure}{L}{0.2\textwidth}
    \centering
    \includegraphics[width=0.2\textwidth]{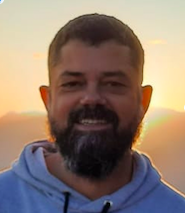}
\end{wrapfigure}

\noindent \textbf{João Terêncio Dias} was born in Tarumirim, Minas Gerais, Brazil, on April 10, 1971. Graduated in Telecommunications Engineering from the Fluminense Federal University (UFF), in 2002, MSc in Electrical Engineering (Digital Communication Systems) from the Military Engineering Institute (IME), in 2006, PhD in Electrical Engineering (Digital Signal Processing) from the Federal University of Rio de Janeiro (UFRJ), in 2013, and post-doctorate in Electrical Engineering (Digital Signal Processing Applied to Communications Systems) at Pontifical Catholic University of Rio de Janeiro (PUC-Rio) in 2016, all in Rio de Janeiro, Brazil. He joined the teaching staff of the Rio de Janeiro Technical School Support Foundation (FAETEC) in 2002, where he worked as a Telecommunications Professor until 2008, when he transferred to the Federal Center for Technological Education of Rio de Janeiro (CEFET/RJ), where he is now Full Professor. Between 2006 and 2008 he was a Substitute Professor at the Faculty of Engineering at the State University of Rio de Janeido (UERJ) and in January 2017 he was a visiting researcher at the University of York, in England. He prepared and reviewed questions for the National Student Performance Exam (ENADE) for Engineering and Computing. He was also a member of the Mathematics team for preparing and reviewing the National High School Exam (ENEM). He is currently leader of the Quantum Communication Research Group (CQRG) and develops extension and research projects in Telecommunications and Computing. His research interests include low-resolution converters, quantum algorithms, wireless communication systems, communication theory, and digital signal processing. João has been a member of the Brazilian Telecommunications Society (SBrT) since 2003 and a member of the Institute of Electrical and Electronic Engineers (IEEE). \newline

\begin{wrapfigure}{L}{0.2\textwidth}
    \centering
    \includegraphics[width=0.2\textwidth]{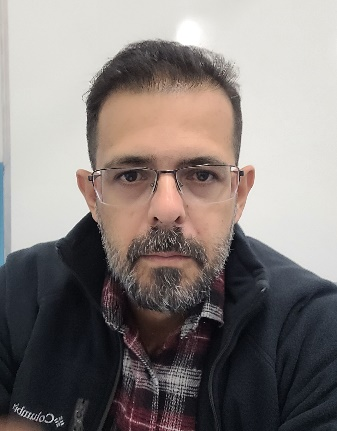}
\end{wrapfigure}

\vfill

\noindent \textbf{Demerson Nunes Gonçalves} is a professor in the Mathematics Department at CEFET/RJ since 2015. He holds a bachelor’s degree in Mathematics Education from the Federal University of Espírito Santo (2002), a master's degree in Computational Modeling from the National Laboratory for Scientific Computing (2005), and a doctorate in Computational Modeling from the National Laboratory for Scientific Computing (2009). His research interests lie at the intersection of mathematics, scientific computing and quantum physics, with a focus on both practical and theoretical applications. His work involves the development of advanced methods and algorithms to solve complex problems in the areas of computational group theory, quantum algorithms, quantum error-correcting codes, and quantum machine learning.

\end{document}